\documentclass[journal]{IEEEtran}

\usepackage{cite}
\usepackage{graphicx}
\usepackage{amsmath}
\usepackage{amsfonts}
\usepackage{amssymb}
\usepackage{mathrsfs}
\newtheorem{theorem}{Theorem}
\newtheorem{definition}{Definition}
\newtheorem{proposition}{Proposition}

\begin{document}

\title{Feedback Control of Quantum State Reduction}
\author{Ramon~van~Handel,~John~K.~Stockton,~and~Hideo~Mabuchi%
\thanks{This work was supported by the ARO (DAAD19-03-1-0073) and the 
Caltech MURI Center for Quantum Networks (DAAD19-00-1-0374).  JKS 
acknowledges a Hertz fellowship.}%
\thanks{The authors are with the departments of Physics and Control \& 
Dynamical Systems, California Institute of Technology 266-33, Pasadena, CA 
91125 USA (e-mail: ramon@its.caltech.edu).}%
}
\markboth{\tiny This work has been submitted to the IEEE for possible 
publication. Copyright may be transferred without notice, after which this 
version may no longer be accessible.}{}
\pubid{~}
\maketitle

\begin{abstract}
Feedback control of quantum mechanical systems must take into account the 
probabilistic nature of quantum measurement. We formulate quantum feedback 
control as a problem of stochastic nonlinear control by considering 
separately a quantum filtering problem and a state feedback control 
problem for the filter.  We explore the use of stochastic Lyapunov 
techniques for the design of feedback controllers for quantum spin systems 
and demonstrate the possibility of stabilizing one outcome of a quantum 
measurement with unit probability.
\end{abstract}


\section{Introduction}
\PARstart{I}{t is} a basic fact of nature that at small scales---at the 
level of atoms and photons---observations are inherently probabilistic,
as described by the theory of quantum mechanics.  The traditional 
formulation of quantum mechanics is very different, however, from the way 
stochastic processes are modeled.  The theory of quantum measurement is 
notoriously strange in that it does not allow all quantum observables to 
be measured simultaneously.  As such there is yet much progress to be made 
in the extension of control theory, particularly feedback control, to the 
quantum domain.

One approach to quantum feedback control is to circumvent measurement
entirely by directly feeding back the physical output from the system 
\cite{s:alloptical,s:ya2}.  In quantum optics, where the system is 
observed by coupling it to a mode of the electromagnetic field, this 
corresponds to all-optical feedback.  Though this is in many ways an 
attractive option it is clear that performing a measurement allows greater 
flexibility in the control design, enabling the use of sophisticated 
in-loop signal processing and non-optical feedback actuators.  Moreover, 
it is known that some quantum states obtained by measurement are not 
easily prepared in other ways \cite{s:stockton,s:geremia,s:knill}.

We take a different route to quantum feedback control, where measurements 
play a central role.  The key to this approach is that quantum theory, 
despite its entirely different appearance, is in fact very closely 
related to Kolmogorov's classical theory of probability.  The essential 
departure from classical probability is the fact that in quantum theory 
observables need not commute, which precludes their simultaneous 
measurement.  Kolmogorov's theory is not equipped to deal with such 
objects: one can always obtain a joint probability distribution for random 
variables on a probability space, implying that they can be measured 
simultaneously.  Formalizing these ideas leads naturally to the rich field 
of {\it noncommutative} or {\it quantum probability} 
\cite{s:maassen,s:biane,s:meyer}.  Classical probability is obtained as a 
special case if we consider only commuting observables.

Let us briefly recall the setting of stochastic control theory.  The 
system dynamics and the observation process are usually described by 
stochastic differential equations of the It\^o type.  A generic approach 
to stochastic control \cite{s:mortensen,s:astrom} separates the problem 
into two parts.  First one constructs a filter which propagates our 
knowledge of the system state given all observations up to the current 
time.  Then one finds a state feedback law to control the filtering 
equation.  Stochastic control theory has traditionally focused on linear 
systems, where the optimal (LQG) control problem can be solved explicitly.

A theory of quantum feedback control with measurement can now be developed
simply by replacing each ingredient of stochastic control theory by its
noncommutative counterpart.  In this framework, the system and 
observations are described by quantum stochastic differential equations.  
The next step is to obtain quantum filtering equations 
\cite{s:belavkin,s:belavkz1,s:belavkz2,s:bouten}.  Remarkably, the filter 
is a classical It\^o equation due to the fact that the output signal of a 
laboratory measuring device is a classical stochastic process.  The 
remaining control problem now reduces to a problem of classical stochastic 
nonlinear control.  As in the classical case, the optimal control problem 
can be solved explicitly for quantum systems with linear dynamics.

\pubidadjcol

The field of quantum stochastic control was pioneered by V.\,P.\ Belavkin 
in a remarkable series of papers 
\cite{s:belavkin2,s:belavkin,s:belavkz1,s:belavkz2} in which the quantum 
counterparts of nonlinear filtering and LQG control were developed.
The advantage of the quantum stochastic approach is that the details of 
quantum probability and measurement are hidden in a quantum filtering 
equation and we can concentrate our efforts on the classical control 
problem associated with this equation.  Recently the quantum filtering 
problem was reconsidered by Bouten {\it et al.}\ \cite{s:bouten} and 
quantum 
optimal control has received some attention in the physics literature 
\cite{s:doherty1,s:doherty2}.

The goal of this paper is twofold.  We review the basic ingredients of 
quantum stochastic control: quantum probability, filtering, and the 
associated geometric structures.  We then demonstrate the use of this 
framework in a {\it nonlinear} control problem.  To this end, we study in 
detail an example directly related to our experimental apparatus 
\cite{s:geremia}.  As this is not a linear system, the optimal control 
problem is intractable and we must resort to methods of stochastic 
nonlinear control.  We use stochastic Lyapunov techniques to design 
stabilizing controllers, demonstrating the feasibility of such an 
approach.

We are motivated in studying the quantum control problem by recent 
developments in experimental quantum optics \cite{s:geremia,s:exp1,s:exp2,s:exp3}.
Technology has now matured to the point that state-of-the-art experiments 
can monitor and manipulate atomic and optical systems in real time {\it at 
the quantum limit;} i.e.\ the sources of extraneous noise are sufficiently 
suppressed that essentially all the noise is fundamental in nature.  The 
experimental implementation of quantum control systems is thus within 
reach of current experiments, with important applications in e.g.\ 
precision metrology \cite{s:exp3,s:metrol1,s:metrol2,s:metrol3} and 
quantum computing \cite{s:qcomp1,s:qcomp2}.  Further development of 
quantum control theory is an essential step in this direction.

This paper is organized as follows: in section \ref{sec:quantum} we give 
an introduction to quantum probability and sketch a simple derivation of 
quantum filtering equations.  We also introduce the particular physical 
system that we study in the remainder of the paper.  In section 
\ref{sec:systems} we study the dynamical behavior of the filtering 
equation and the underlying geometric structures.  Finally, section 
\ref{sec:spin} is devoted to the design of stabilizing controllers using
stochastic Lyapunov methods.

\section{Quantum probability and filtering}
\label{sec:quantum}

The purpose of this section is to clarify the connections between quantum 
mechanics and classical probability theory.  The emphasis is not on rigor 
as we aim for a brief but broad overview; we refer to the references for a 
complete treatment.

\subsection{Finite-dimensional quantum probability}
\label{sec:finite}

We begin by reviewing some of the traditional elements of quantum 
mechanics (e.g.\ \cite{s:merzbacher}) with a probabilistic flavor.

An {\it observable} of a finite-dimensional quantum system is represented 
by a self-adjoint linear operator $X=X^*$ on some underlying 
finite-dimensional complex Hilbert space $\mathcal{H}$
($*$ denotes Hermitian conjugation).  Every self-adjoint operator has a 
spectral decomposition
\begin{equation}
\label{eq:specdec}
	X=\sum_i\lambda_iP_i,
	~~~~~~~\lambda_i\in\mathbb{R},~~P_i=P_i^2=P_i^*
\end{equation}
where $\lambda_i$ are the eigenvalues of $X$ and $P_i$ are projectors onto 
orthogonal eigenspaces in $\mathcal{H}$ such that $\sum_iP_i={\rm 
Id}_\mathcal{H}$.

If we were to measure $X$ we would obtain one of the values $\lambda_i$ 
as the measurement outcome.  The $P_i$ represent the events that can be
measured.  To complete the picture we still need a probability measure.  
This is provided by the {\it density operator} 
$\rho$, which is a linear operator on $\mathcal{H}$ satisfying
\begin{equation}
	\rho=\rho^*,~~~~~~~{\rm Tr}\rho=1,~~~~~~~\rho\ge 0.
\end{equation}
The probability of an event $P_i$ is given by
\begin{equation}
\label{eq:collprodir}
	p_i={\rm Tr}[\rho P_i].
\end{equation}
We can now easily find the expectation of $X$:
\begin{equation}
	\langle X\rangle=\sum_i\lambda_i\,{\rm Tr}[\rho P_i]=
		{\rm Tr}[\rho X].
\end{equation}
In quantum mechanics $\rho$ is also called the system state.

As in classical probability, it will be useful to formalize these ideas 
into a mathematical theory of quantum probability 
\cite{s:maassen,s:biane,s:meyer}.  The main ingredient of the theory is 
the quantum probability space $(\mathcal{A},\rho)$.  Here $\mathcal{A}$ is 
a $*$-algebra, i.e.\ an algebra with involution $*$ of linear 
operators on $\mathcal{H}$, and $\rho$ is the associated state.  An 
observable on $(\mathcal{A},\rho)$ is a sum of the form (\ref{eq:specdec}) 
with $P_i\in\mathcal{A}$.  In the finite-dimensional case this implies 
that every observable is a member of $\mathcal{A}$, but we will see that 
this need not be the case in infinite dimensions.

$\mathcal{A}$ does not necessarily contain all self-adjoint operators on 
$\mathcal{H}$.  Of special importance is the case in which $\mathcal{A}$ 
is a commutative algebra, i.e.\ all the elements of $\mathcal{A}$ commute 
($[X,Y]=XY-YX=0$ $\forall X,Y\in\mathcal{A}$.)  It is easily verified that 
there is a one-to-one correspondence (up to isomorphism) between 
commutative quantum probability spaces $(\mathcal{A},\rho)$ and classical 
probability spaces $(\Omega,\mathcal{F},\mathbb{P})$ with ${\rm 
card}\,\Omega=\dim\mathcal{H}$.  As $\mathcal{A}$ is commutative we may 
represent all its elements by diagonal matrices; the diagonals are then 
interpreted as functions $f:\Omega\to\mathbb{R}$.  The projectors 
$P_i\in\mathcal{A}$ now correspond to indicator functions $\chi_{A_i}$ on 
$\Omega$ and hence define the $\sigma$-algebra $\mathcal{F}=\{A_i\}$.  
Finally, $\mathbb{P}$ is defined by $\mathbb{P}[A_i]={\rm Tr}[P_i\rho]$.

Clearly classical probability is a special case of quantum probability.
However, noncommutative $\mathcal{A}$ are inherent to quantum mechanical 
models.  Suppose $A,B$ are two events (projectors) that do not commute.  
Then $A$ and $B$ cannot be diagonalized simultaneously, and hence they 
cannot be represented as events on a single classical probability space.  
Suppose we wish to measure $A$ and $B$ simultaneously, i.e.\ we ask what 
is the probability of the event ($A$ and $B$)?  In the classical case this 
would be given by the joint probability $\mathbb{P}[A,B]=\mathbb{P}[A\cap 
B]=\mathbb{E}[\chi_A\chi_B]$.  However in the noncommutative case this 
expression is ambiguous as ${\rm Tr}[\rho AB]\ne{\rm Tr}[\rho BA]$.
We conclude that ($A$ and $B$) is an invalid question and its probability 
is undefined.  In this case, the events $A$ and $B$ are said to be {\it 
incompatible}.  Similarly, two observables on $\mathcal{A}$ can be 
measured simultaneously only if they commute.

We conclude this section with the important topic of conditional 
expectation.  A traditional element of the theory of quantum measurement 
is the projection postulate, which can be stated as follows.  Suppose we 
measure an observable $X$ and obtain the outcome $\lambda_i$.  Then the 
measurement causes the state to collapse:
\begin{equation}
\label{eq:collapse}
	\rho|_i=\frac{P_i\rho P_i}{{\rm Tr}[\rho P_i]}.
\end{equation}
Suppose that we measure another observable $X'$ after measuring $X$.  
Using (\ref{eq:collapse}) we write
\begin{equation}
\label{eq:quantco}
	P[X'=\lambda'_j|X=\lambda_i]=
		{\rm Tr}[P'_j\rho|_i]=
		\frac{{\rm Tr}[\rho P_iP'_jP_i]}{{\rm Tr}[\rho P_i]}.
\end{equation}
Now compare to the definition of conditional probability in classical 
probability theory:
\begin{equation}
\label{eq:classco}
	\mathbb{P}[B|A]=\frac{\mathbb{P}[B\cap A]}{\mathbb{P}[A]},
	~~~~~~~ A,B\in\mathcal{F}.
\end{equation}
Clearly (\ref{eq:quantco}) and (\ref{eq:classco}) are completely 
equivalent if $X,X'$ commute.  It is now straightforward to define the 
quantum analog of conditional expectation:
\begin{equation}
\label{eq:qcondex}
	\mathcal{E}[X'|\mathcal{B}]=
	\sum_i\frac{{\rm Tr}[\rho P_iX'P_i]}{{\rm Tr}[\rho P_i]}P_i.
\end{equation}
Here $\mathcal{B}$ is the $*$-algebra generated by $X$, i.e.\ it is 
the algebra whose smallest projectors are $P_i$.  This definition also
coincides with the classical conditional expectation if $X,X'$ commute.

We obtain ambiguous results, however, when $X,X'$ do not commute, as then 
the fundamental property 
$\langle\mathcal{E}[X'|\mathcal{B}]\rangle=\langle X'\rangle$
is generally lost.  This implies that if we measure an observable, but 
``throw away'' the measurement outcome, the expectation of the observable 
may change.  Clearly this is inconsistent with the concept of conditional 
expectation which only changes the observer's state of knowledge about 
the system, but this is not surprising: noncommuting $X,X'$ cannot be 
measured simultaneously, so any attempt of statistical inference of $X'$ 
based on a measurement of $X$ is likely to be ambiguous.  To avoid this 
problem we define the conditional expectation only for the case that 
$X'$ commutes with every element of $\mathcal{B}$.  The measurement 
$\mathcal{B}$ is then said to be {\it nondemolition} \cite{s:belavkin} 
with respect to $X'$.  

The essence of the formalism we have outlined is that the foundation of 
quantum theory is an extension of classical probability theory.  This 
point of view lies at the heart of quantum stochastic control.  The 
traditional formulation of quantum mechanics can be directly recovered 
from this formalism.  Even the nondemolition requirement is not a 
restriction: we will show that the collapse rule (\ref{eq:collapse}) 
emerges in a quantum filtering theory that is based entirely on 
nondemolition measurements.

\subsection{Infinite-dimensional quantum probability}

The theory of the previous section exhibits the main features of quantum 
probability, but only allows for finite-state random variables.  A general 
theory which allows for continuous random variables is developed along 
essentially the same lines where linear algebra, the foundation of 
finite-dimensional quantum mechanics, is replaced by functional analysis.  
We will only briefly sketch the constructions here; a lucid introduction 
to the general theory can be found in \cite{s:maassen}.

A quantum probability space $(\mathcal{A},\rho)$ consists of a Von Neumann 
algebra $\mathcal{A}$ and a state $\rho$.  A Von Neumann algebra is a
$*$-algebra of bounded linear operators on a complex Hilbert space 
$\mathcal{H}$ and $\rho:\mathcal{A}\to\mathbb{C}$ is a linear map 
such that $\rho({\rm Id}_\mathcal{H})=1$, $\rho(A^* 
A)\ge 0$ $\forall A\in\mathcal{A}$ and $\rho(A^*A)=0$ iff $A=0$.
We gloss over additional requirements related to limits of sequences of 
operators.  It is easily verified that the definition reduces in the 
finite-dimensional case to the theory in the previous section, where the 
density operator $\rho$ is identified with the map $X\mapsto{\rm Tr}[\rho X]$.
We always assume ${\rm Id}_\mathcal{H}\in\mathcal{A}$.

As in the finite-dimensional case there is a correspondence between 
classical probability spaces and commutative algebras.  Given the 
classical space $(\Omega,\mathcal{F},\mathbb{P})$ the associated quantum 
probability space is constructed as follows:
\begin{equation}
	\mathcal{H}=L^2(\Omega;\mathbb{C}),~~~~~
	\mathcal{A}=L^\infty(\Omega;\mathbb{C}),~~~~~
	\rho:f\mapsto\int_\Omega f\,d\mathbb{P}
\end{equation}
where $\mathcal{A}$ acts on $\mathcal{H}$ by pointwise multiplication.  
Conversely, every commutative quantum probability space corresponds to a 
classical probability space.  This fundamental result in the theory of 
operator algebras is known as Gel'fand's theorem.

Observables are represented by linear operators that are self-adjoint with 
respect to some dense domain of $\mathcal{H}$.  The spectral decomposition 
(\ref{eq:specdec}) is now replaced by the spectral theorem of functional 
analysis, which states that every self-adjoint operator $X$ can be 
represented as
\begin{equation}
	X=\int_\mathbb{R}\lambda\,E(d\lambda),~~~~~~~
		E:\mathcal{B}_\mathbb{R}\to\mathcal{P}(\mathcal{H}).
\end{equation}
Here $E$ is the spectral or projection-valued measure associated to $X$, 
$\mathcal{P}(\mathcal{H})$ is the set of all projection operators on 
$\mathcal{H}$, and $\mathcal{B}_\mathbb{R}$ is the Borel $\sigma$-algebra 
on $\mathbb{R}$.  $X$ is {\it affiliated} to $\mathcal{A}$ if
$E(\Lambda)\in\mathcal{A}$ $\forall\Lambda\in\mathcal{B}_\mathbb{R}$,
replacing the concept of measurability in classical probability theory.  
For $X$ affiliated to $\mathcal{A}$ the probability law and expectation 
are given by
\begin{equation}
	P[X\in\Lambda]=\rho(E(\Lambda)),~~~~~~~
	\langle X\rangle=\int_\mathbb{R}\lambda\,\rho(E(d\lambda)).
\end{equation}
Note that unlike in finite dimensions not all observables affiliated to 
$\mathcal{A}$ are elements of $\mathcal{A}$; observables may be unbounded 
operators, while $\mathcal{A}$ only contains bounded operators.

It remains to generalize conditional expectations to the 
infinite-dimensional setting, a task that is not entirely straightforward 
even in the classical case.  Let $\mathcal{B}\subset\mathcal{A}$ be a 
commutative Von Neumann subalgebra.  As before we will only define 
conditional expectations for observables that are not demolished by 
$\mathcal{B}$, i.e.\ for observables affiliated to the commutant 
$\mathcal{B}'=\{A\in\mathcal{A}:[A,B]=0~\forall B\in\mathcal{B}\}$.
\begin{definition}
\label{def:condex}
	The conditional expectation onto $\mathcal{B}$ is the
	linear surjective map $\mathcal{E}[\cdot|\mathcal{B}]:\mathcal{B}'
	\to\mathcal{B}$ with the following properties: for all
	$A\in\mathcal{B}'$
	\begin{enumerate}
	\item $\mathcal{E}[{\rm Id}_\mathcal{H}|\mathcal{B}]
		={\rm Id}_\mathcal{H}$,
	\item $\mathcal{E}[A|\mathcal{B}]\ge 0$ if $A\ge 0$,
	\item $\mathcal{E}[B_1AB_2|\mathcal{B}]=
			B_1\mathcal{E}[A|\mathcal{B}]B_2$
		$\forall B_1,B_2\in\mathcal{B}$, and
	\item $\rho(\mathcal{E}[A|\mathcal{B}])=\rho(A)$.
	\end{enumerate}
	The definition extends to any observable $X$ affiliated to 
	$\mathcal{B}'$ by operating $\mathcal{E}[\cdot|\mathcal{B}]$ on 
	the associated spectral measure.
\end{definition}

It is possible to prove (e.g.\ \cite{s:bouten}) that the conditional 
expectation exists and is unique.

\subsection{Quantum stochastic calculus}

Having extended probability theory to the quantum setting, we now 
sketch the development of a quantum It\^o calculus.

We must first find a quantum analog of the Wiener process.  Denote by 
$(\Omega,\mathcal{F},\mathbb{P})$ the canonical Wiener space of a 
classical Brownian motion.  The analysis in the previous section suggests 
that quantum Brownian motion will be represented by a set of observables 
on the Hilbert space $\Gamma=L^2(\Omega;\mathbb{C})$.  Define the 
symmetric Fock space over $L^2(U)$ as
\begin{equation}
\label{eq:fock}
	\Gamma_s(L^2(U))=\mathbb{C}\oplus\bigoplus_{n=1}^\infty
		L^2(U;\mathbb{C})^{\odot n},~~~~~~~
	U\subset\mathbb{R}_+
\end{equation}
where $\odot$ denotes the symmetrized tensor product.  It is well known 
in stochastic analysis (e.g.\ \cite{s:meyer}) that $\Gamma$ and 
$\Gamma_s(L^2(\mathbb{R}_+))$ are isomorphic, as every $L^2$-functional 
on $\Omega$ is associated to its Wiener chaos expansion.  Now define the 
operators
\begin{equation}
\begin{split}
	A_g{\bf k} &=
		\sum_{i=1}^n k_1\odot\cdots\odot\hat k_i\odot\cdots\odot 
			k_n \int_{\mathbb{R}_+} g^*k_i\,dt \\
	A^*_g{\bf k} &=
		g\odot k_1\odot\cdots\odot k_n
\end{split}
\end{equation}
where ${\bf k}=k_1\odot\cdots\odot k_n$, $g,k_i\in L^2(\mathbb{R}_+)$ and 
$\hat k_i$ means that the term $i$ is omitted.  It is sufficient to define 
the operators for such vectors as their linear span $\Gamma_0$ is dense in 
$\Gamma$.  We get
\begin{equation}
\label{eq:fieldcommut}
	[A_g,A_h]=[A_g^*,A_h^*]=0,~~~~~~~
	[A_g,A_h^*]=\int_{\mathbb{R}_+} g^*h\,dt
\end{equation}
and indeed $(v,A_gw)=(A_g^* v,w)$ for $v,w\in\Gamma_0$.

We will construct Wiener processes from $A$ and $A^*$, but first we 
must set up the quantum probability space.  We take $\mathcal{A}$ to 
contain all bounded linear operators on $\Gamma$.  To construct $\rho$ 
consider the vector $\Delta=1\oplus 0\in\Gamma_s(L^2(\mathbb{R}_+))$. Then
\begin{equation}
\label{eq:vacuum}
	\rho:\mathcal{A}\to\mathbb{C},~~~~
	\rho(X)=(\Delta,X\Delta).
\end{equation}
Now consider the operator $A_g^*+A_{g}$.  Using (\ref{eq:fieldcommut})
and the Baker-Campbell-Hausdorff lemma we obtain
\begin{equation}
	\langle e^{i(A_g^*+A_{g})}\rangle=
	(\Delta,e^{iA_g^*}e^{-\frac{1}{2}\|g\|^2} 
		e^{iA_{g}}\Delta)=
	e^{-\frac{1}{2}\|g\|^2}
\end{equation}
where $\|g\|^2$ is the integral of $|g|^2$ over $\mathbb{R}_+$.  However, 
the characteristic functional of a classical Wiener process is
\begin{equation}
	\mathbb{E}[e^{i\int_0^\infty\!\tilde 
		g(t)dW_t}]=e^{-\frac{1}{2}\|\tilde g\|^2}
\end{equation}
where $\tilde g$ is a real function.  Clearly $A_g^*+A_{g}$ is 
equivalent in law to a classical Wiener integral, and any $Q_t=
A_{g_t}^*+A_{g_t}$ with $g_t(s)=\chi_{[0,t]}(s)\,e^{i\varphi(s)}$ is a 
quantum Wiener process.  

It is easy to verify that $[Q_t,Q_s]=0$ $\forall t,s$.  This important 
property allows us to represent all $Q_t$, $t\in\mathbb{R}_+$ on a single 
classical probability space, and hence $Q_t$ is entirely equivalent to a 
classical Wiener process.  Two such processes with different $\varphi$ do 
not commute, however, and are thus incompatible.

The Fock space (\ref{eq:fock}) has the following factorization property: 
for any sequence of times $t_1<t_2<\ldots<t_n\in\mathbb{R}_+$
\begin{equation}
	\Gamma=\Gamma_{t_1]}\otimes
	\Gamma_{t_1,t_2}\otimes\Gamma_{t_2,t_3}\otimes\cdots
	\otimes\Gamma_{t_{n-1},t_n}\otimes\Gamma_{[t_n}
\end{equation}
with $\Gamma_{s,t}=\Gamma_s(L^2([s,t]))$, $\Gamma_{t]}=\Gamma_{0,t}$ and 
$\Gamma_{[t}=\Gamma_{t,\infty}$.  Thus $\Gamma$ can be formally considered 
as a continuous tensor product over $\Gamma_s(L^2(\{t\}))$, a construction 
often used implicitly in physics literature.  A process $S_t$ is 
called {\it adapted} if $S_t=S_{t]}\otimes{\rm Id}$ in 
$\Gamma_{t]}\otimes\Gamma_{[t}$ for every $t\in\mathbb{R}_+$.  $Q_t$ is 
adapted for any $\varphi$.

It is customary to define the standard noises
\begin{equation}
	A_t=A_{\chi_{[0,t]}},~~~~~~~
	A_t^*=A_{\chi_{[0,t]}}^*,~~~~~~~t\in\mathbb{R}_+.
\end{equation}
One can now define It\^o integrals and calculus with respect to 
$A_t,A_t^*$ in complete analogy to the classical case.  We will only 
describe the main results, due to Hudson and Parthasarathy 
\cite{s:hudpar}, and refer to \cite{s:biane,s:meyer,s:hudpar} for the full 
theory.

Let $\mathcal{H}$ be the Hilbert space of the system of interest; we will 
assume that $\dim\mathcal{H}<\infty$.  Now let $\mathcal{A}$ be the set 
of all bounded operators on $\mathcal{H}\otimes\Gamma$.  The state
$\rho=\rho_\mathcal{H}\otimes\rho_\Gamma$ is given in terms some state 
$\rho_\mathcal{H}$ on $\mathcal{H}$ and $\rho_\Gamma$ as defined in 
(\ref{eq:vacuum}).  The Hudson-Parthasarathy equation
\begin{equation}
\label{eq:hudpar}
	U_{s,t}={\rm Id}+\int_s^t(L\,dA_t^*-L^*\,dA_t
		-(iH+\tfrac{1}{2}L^* L)\,dt)\,U_{s,t}
\end{equation}
defines the flow $U_{s,t}$ of the noisy dynamics.  Here $L$ and $H$ are 
operators of the form $L\otimes{\rm Id}$ on $\mathcal{H}\otimes\Gamma$ and 
$H$ is self-adjoint.  It can be shown that $U_{s,t}$ is a unitary 
transformation of $\mathcal{H}\otimes\Gamma_{s,t}$ and 
$U_{s,t}=U_{k,t}U_{s,k}$.  Given an observable $S$ at time $0$, the 
flow defines the associated process $S_t=U_{0,t}^* S U_{0,t}$.

Quantum stochastic differential equations are easily manipulated using the 
following rules.  The expectation of any integral over $dA_t$ or 
$dA_t^*$ vanishes.  The differentials $dA_t,dA_t^*$ commute with any 
adapted process.  Finally, the quantum It\^o rules are 
$dA_t\,dA_t^*=dt$, $dA_t^2=(dA_t^*)^2=dA_t^*\,dA_t=0$.

Let $X\in\mathcal{H}$ be any system observable; its time evolution is 
given by $j_t(X)=U_{0,t}^*(X\otimes{\rm Id})U_{0,t}$. 
We easily obtain
\begin{equation}
\label{eq:qito}
	dj_t(X)=j_t(\mathcal{L}X)\,dt
		+j_t([L^*,X])\,dA_t
		+j_t([X,L])\,dA_t^*
\end{equation}
where $\mathcal{L}X=i[H,X]+L^* X L-\tfrac{1}{2}(L^* LX+XL^* L)$.  
This expression is the quantum analog of the classical It\^o formula
\begin{equation}
\label{eq:cito}
	dj_t(f)=j_t(\mathscr{L}f)\,dt
		+j_t(\Sigma f)\,dW_t
\end{equation}
where $j_t(f)=f(x_t)$ with $dx_t=b(x_t)\,dt+\sigma(x_t)\,dW_t$, 
$\mathscr{L}$ is the infinitesimal generator of $x_t$ and $\Sigma 
f=\sigma^i\partial_if$.  Similarly, $\mathcal{L}$ is called the generator 
of the quantum diffusion $U_{s,t}$.

In fact, the quantum theory is very similar to the classical theory of 
stochastic flows \cite{s:bismut,s:arnold2} with one notable exception: the 
existence of incompatible observables does not allow for a unique sample 
path interpretation ($x_t$ in the classical case) of the underlying 
system.  Hence the dynamics is necessarily expressed in terms of 
observables, as in (\ref{eq:qito}).

\subsection{Measurements and filtering}

We now complete the picture by introducing observations and conditioning 
the system observables on the observed process.  The following treatment 
is inspired by \cite{s:belavkz1,s:belavkz2}.

\subsubsection{Classical filtering}

To set the stage for the quantum filtering problem we first treat its
classical counterpart.  Suppose the system dynamics (\ref{eq:cito}) 
is observed as $y_t$ with
\begin{equation}
\label{eq:classobs}
	dy_t=j_t(h)\,dt+\kappa\,dV_t
\end{equation}
for uncorrelated noise $V_t$ with strength $\kappa>0$.  We wish to 
calculate the conditional expectation 
$\pi_t(f)=\mathbb{E}[j_t(f)|\mathcal{F}_t^y]$.

Recall the classical definition: $\mathbb{E}[X|\mathcal{F}]$ is the 
$\mathcal{F}$-measurable random variable such that 
$\mathbb{E}[\mathbb{E}[X|\mathcal{F}]Y]=\mathbb{E}[XY]$ for all 
$\mathcal{F}$-measurable $Y$.  Suppose $\mathcal{F}$ is generated by some 
random variable $F$.  The definition suggests that to prove
$\hat X=\mathbb{E}[X|\mathcal{F}]$ for some $\mathcal{F}$-measurable 
$\hat X$, it should be sufficient to show that
\begin{equation}
	\mathbb{E}[\hat Xe^{F\xi}]=\mathbb{E}[Xe^{F\xi}]
	~~~~~~~\forall \xi\in\mathbb{R},
\end{equation}
i.e., the conditional generating functions coincide.

We will apply this strategy in the continuous case.  As $\pi_t(f)$ is an
$\mathcal{F}_t^y$-semimartingale we introduce the ansatz
\begin{equation}
	d\pi_t(f)=C_t\,dt+D_t\,dy_t
\end{equation}
with $C_t,D_t$ $\mathcal{F}_t^y$-adapted.  We will choose $C_t,D_t$ such 
that $\mathbb{E}[e_t^g\pi_t(f)]=\mathbb{E}[e_t^gj_t(f)]$ for all functions 
$g$, where
\begin{equation}
	e_t^g=e^{\int_0^tg(s)dy_s-\tfrac{1}{2}\kappa^2\int_0^tg(s)^2ds},
	~~~~~~~ de_t^g=g(t)e_t^g\,dy_t.
\end{equation}
The It\^o correction term in the exponent was chosen for convenience and 
does not otherwise affect the procedure.

Using It\^o's rule and the usual properties of conditional 
expectations we easily obtain
\begin{align}
	\frac{d\,\mathbb{E}[e_t^gj_t(f)]}{dt} &=
	\mathbb{E}[e_t^g\pi_t(\mathscr{L}f)+g(t)e_t^g\pi_t(hf)] \\
\begin{split}
	\frac{d\,\mathbb{E}[e_t^g\pi_t(f)]}{dt} &=
	\mathbb{E}[e_t^g(C_t+\pi_t(h)D_t) \\
		&\phantom{=\mathbb{E}[}
		+g(t)e_t^g(\kappa^2D_t+\pi_t(h)\pi_t(f))].
\end{split}
\end{align}
Requiring these expressions to be identical for any $g$ gives
\begin{equation}
\label{eq:kushners}
	d\pi_t(f)=\pi_t(\mathscr{L}f)\,dt+\kappa^{-1}
		(\pi_t(hf)-\pi_t(h)\pi_t(f))\,d\overline{W}_t
\end{equation}
where the innovations process $d\overline{W}_t=\kappa^{-1}(dy_t-\pi_t(h)\,dt)$
is a Wiener process.  Eq.\ (\ref{eq:kushners}) is the well-known 
Kushner-Stratonovich equation of nonlinear filtering
\cite{s:hazewinkel,s:liptser}.

\subsubsection{Quantum filtering}

The classical approach generalizes directly to the quantum case.  The main 
difficulty here is how to define in a sensible way the observation 
equation (\ref{eq:classobs})?

We approach the problem from a physical perspective \cite{s:barchielli}.
The quantum noise represents an electromagnetic field coupled to the 
system (e.g.\ an atom.)  Unlike classically, where any observation is in 
principle admissible, a physical measurement is performed by placing a 
detector in the field.  Hence the same noise that drives the system is 
used for detection, placing a physical restriction on the form of the 
observation.

We will consider the observation 
$Y_t'=U_{0,t}^*(A_t^*+A_t)U_{0,t}+\kappa(B_t^*+B_t)$.  Here $B_t$ 
is a noise uncorrelated from $A_t$ that does not interact with the system 
(the Hilbert space is $\mathcal{H}\otimes\Gamma\otimes\Gamma$, etc.)  
Physically we are measuring the field observable $A_t^*+A_t$ after 
interaction with the system, corrupted by uncorrelated noise of strength 
$\kappa>0$.  Using the It\^o rule and (\ref{eq:hudpar}) we get
\begin{equation}
	dY_t'=j_t(L^*+L)\,dt+dA_t^*+dA_t+\kappa(dB_t^*+dB_t).
\end{equation}
It is customary in physics to use a normalized observation $Y_t$ such that 
$dY_t^2=dt$.  We will use the standard notation
\begin{equation}
\label{eq:measmeas}
	dY_t=\sqrt{\eta}\,(j_t(L^*+L)\,dt+dA_t^*+dA_t)
		+\sqrt{1-\eta}\,dV_t
\end{equation}
where $V_t=B_t^*+B_t$ and $\eta=(1+\kappa^2)^{-1}\in(0,1]$.

$Y_t'$ and $Y_t$ satisfy the following two crucial properties:
\begin{enumerate}
\item
$Y_t'$ is {\it self-nondemolition}, i.e.\ $[Y_t',Y_s']=0$ $\forall s<t$. 
To see this, note that $[Y_t',Y_s']=[U_{0,t}^* Q_tU_{0,t}, U_{0,s}^* 
Q_sU_{0,s}]$ with $Q_t=A_t^*+A_t$. But  $U_{s,t}$ is a unitary 
transformation of $\mathcal{H}\otimes\Gamma_{s,t}$ and $Q_s={\rm 
Id}\otimes Q_{s]}\otimes{\rm Id}$ on 
$\mathcal{H}\otimes\Gamma_{s]}\otimes\Gamma_{[s}$; 
thus we get $U_{s,t}^* Q_sU_{s,t}=Q_sU_{s,t}^* U_{s,t}=Q_s$, so 
$U_{0,s}^* Q_sU_{0,s}=U_{0,t}^* Q_sU_{0,t}$. But then 
$[Y_t',Y_s']=U_{0,t}^*[Q_t,Q_s]U_{0,t}=0$ as we have already seen that 
$Q_t$ is self-nondemolition.
\item
$Y_t'$ is {\it nondemolition}, i.e.\ $[j_t(X),Y_s']=0$ $\forall s<t$ for 
all system observables $X$ on $\mathcal{H}$.  The proof is identical to 
the proof of the self-nondemolition property.
\end{enumerate}
These properties are essential in any sensible quantum filtering theory:
self-nondemolition implies that the observation is a classical stochastic 
process, whereas nondemolition is required for the conditional 
expectations to exist.  A general filtering theory can be developed that 
allows any such observation \cite{s:belavkin,s:belavkz1}; we will restrict 
ourselves to our physically motivated $Y_t$.

We wish to calculate $\pi_t(X)=\mathcal{E}[j_t(X)|\mathcal{B}_t]$ where 
$\mathcal{B}_t$ is the algebra generated by $Y_{s\le t}$.  Introduce the 
ansatz
\begin{equation}
	d\pi_t(X)=C_t\,dt+D_t\,dY_t
\end{equation}
where $C_t,D_t$ are affiliated to $\mathcal{B}_t$.  Define
\begin{equation}
	e_t^g=e^{\int_0^tg(s)dY_s-\tfrac{1}{2}\int_0^tg(s)^2ds},
	~~~~~~~ de_t^g=g(t)e_t^g\,dY_t.
\end{equation}
Using the quantum It\^o rule and Def.\ \ref{def:condex} we get
\begin{align}
\begin{split}
	\frac{d\langle e_t^gj_t(X)\rangle}{dt} &=
	\langle e_t^g\pi_t(\mathcal{L}X)+ \\
		&\phantom{=\langle~~}
		g(t)e_t^g\pi_t(XL+L^* X)\sqrt{\eta}\rangle
\end{split} \\
\begin{split}
	\frac{d\langle e_t^g\pi_t(X)\rangle}{dt} &=
	\langle e_t^g(C_t+\pi_t(L^*+L)D_t\sqrt{\eta})+ \\
		&\phantom{=\langle~~}
		g(t)e_t^g(D_t+\pi_t(L^*+L)\pi_t(X)\sqrt{\eta})\rangle.
\end{split}
\end{align}
Requiring these expressions to be identical for any $g$ gives
\begin{multline}
\label{eq:qfilt}
	d\pi_t(X)=\pi_t(\mathcal{L}X)\,dt
		+\sqrt{\eta}(\pi_t(XL+L^* X) \\
	-\pi_t(L^*+L)\pi_t(X))
		(dY_t-\sqrt{\eta}\,\pi_t(L^*+L)\,dt)
\end{multline}
which is the quantum analog of (\ref{eq:kushners}).  It can be shown that 
the innovations process $dW_t=dY_t-\sqrt{\eta}\,\pi_t(L^*+L)\,dt$ is a 
martingale (e.g.\ \cite{s:bouten}), and hence it is a Wiener process by 
L{\'e}vy's classical theorem.

\subsection{The physical model}

Quantum (or classical) probability does not by itself describe any 
particular physical system; it only provides the mathematical framework in 
which physical systems can be modeled.  The modeling of particular systems 
is largely the physicist's task and a detailed discussion of the issues 
involved is beyond the scope of this article; we limit ourselves to a few 
general remarks.  The main goal of this section is to introduce a 
prototypical quantum system which we will use in the remainder of this 
article.

The emergence of quantum models can be justified in different ways.  The 
traditional approach involves ``quantization'' of classical mechanical 
theories using an empirical quantization rule.  A more fundamental theory 
builds quantum models as ``statistical'' representations of mechanical 
symmetry groups \cite{s:holevo,s:haag}.  Both approaches generally lead to 
the same theory.

\begin{figure}
\centering
\includegraphics[width=3in]{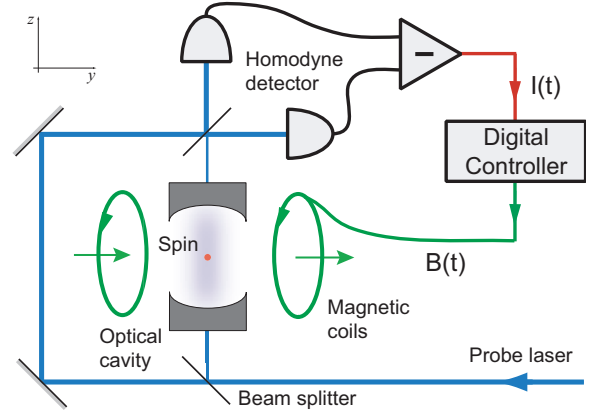}
\caption{Schematic of an experiment for continuous quantum measurement and 
control.  The spin interacts with an optical mode, which is measured 
continuously by homodyne detection.  A magnetic field is used for 
feedback.}
\label{f:schematic}
\end{figure}

The model considered in this paper (Fig.\ \ref{f:schematic}) is 
prototypical for experiments in quantum optics; in fact, it is very 
similar to our laboratory apparatus \cite{s:geremia}.  The system consists 
of a cloud of atoms, collectively labeled ``spin'', interacting with an 
optical field (along $\hat z$) produced by a laser.  After interacting 
with the system the optical field is detected using a photodetector 
configuration known as a homodyne detector.  A pair of magnetic coils
(along $\hat y$) are used as feedback actuators.

The optical and magnetic fields are configured so they only interact, to 
good approximation, with the collective angular momentum degrees of 
freedom of all the atoms \cite{s:dicke}.  Rotational symmetry implies that 
observables of angular momentum must form the rotation Lie algebra 
$\mathfrak{so}(3)$.  If we impose additionally that the total angular 
momentum is conserved, then it is a standard result in quantum mechanics 
\cite{s:merzbacher} that the angular momentum observables form an 
irreducible representation of $\mathfrak{so}(3)$.  Such a system is called 
a spin.

We take $\mathcal{H}$ to be the spin Hilbert space.  Any finite dimension
$2\le\dim\mathcal{H}<\infty$ supports an irrep of $\mathfrak{so}(3)$; the 
choice of $\dim\mathcal{H}=2j+1$ depends on the number of atoms and their 
properties.  We can choose an orthonormal basis 
$\{\psi_m\in\mathcal{H},~m=-j,-j+1,\ldots,j\}$ such that the observables 
$J_{x,y,z}$ of angular momentum around the $x,y,z$-axis are defined 
by\footnote{
	Angular momentum is given in units of
	$\hbar\simeq 1.055 \times 10^{-34}~{\rm kg\,m^2\,s^{-1}}$.
	To simplify the notation we always work in units such that $\hbar=1$.
}
\begin{equation}
\begin{split}
	J_x\psi_m &= c_m\psi_{m+1}+c_{-m}\psi_{m-1} \\
	J_y\psi_m &= ic_m\psi_{m+1}-ic_{-m}\psi_{m-1} \\
	J_z\psi_m &= m\psi_m
\end{split}
\end{equation}
with $c_m=\tfrac{1}{2}\sqrt{(j-m)(j+m+1)}$.  It is easily verified that 
$J_{x,y,z}$ indeed generate $\mathfrak{so}(3)$, e.g.\ $[J_x,J_y]=iJ_z$.

Note that $J_{x,y,z}$ are discrete random variables; the fact that angular 
momentum is ``quantized'', unlike in classical mechanics, is one of the 
remarkable predictions of quantum mechanics that give the theory its name.  
Another remarkable non-classical effect is that $J_{x,y,z}$ are 
incompatible observables.

The noise in our model and its interaction with the atoms emerges 
naturally from quantum electrodynamics, the quantum theory of light 
\cite{s:mandel}.  Physical noise is not white; however, as the correlation 
time of the optical noise is much shorter than the time scale of the spin 
dynamics, a quantum analog of the classical Wong-Zakai procedure 
\cite{s:gough,s:accardi} can be employed to approximate the dynamics by an 
equation of the form (\ref{eq:hudpar}).  In fact, the term 
$-\tfrac{1}{2}L^* L$ in (\ref{eq:hudpar}) is precisely the Wong-Zakai 
correction term that emerges in the white noise limit.

We now state the details of our model without further physical 
justification.  The system is described by (\ref{eq:hudpar}) with 
$L=\sqrt{M}\,J_z$ and $H=B(t)\,J_y$.  Here $M>0$ is the strength of the 
interaction between the light and the atoms; it is regulated 
experimentally by the optical cavity.  $B(t)$ is the applied magnetic 
field and serves as the control input.  Finally, homodyne detection 
\cite{s:barchielli} provides exactly the measurement\footnote{
	In practice one measures not $Y_t$ but its formal derivative 
	$I(t)=dY_t/dt$.  As in classical stochastics we prefer to deal 
	mathematically with the integrated observation $Y_t$ rather than
	the singular ``white noise'' photocurrent $I(t)$.
} (\ref{eq:measmeas}), 
where $\eta$ is determined by the efficiency of the photodetectors.

In the remainder of the paper we will study the spin system of Fig.\ 
\ref{f:schematic}.  Before we devote ourselves entirely to this 
situation, however, we mention a couple of other common scenarios.

Often $L$ is not self-adjoint; in this case, the system can emit or absorb 
energy through interaction with the field.  This situation occurs when the 
optical frequency of the cavity field is resonant with an atomic 
transition.  In our case the frequency is chosen to be far off-resonant; 
this leads to self-adjoint $L$ after adiabatic elimination of the cavity 
dynamics (e.g.\ \cite{s:doherty1}).  The filter dynamics in this scenario, 
to be described below, is known as state reduction.  The sequence of 
approximations that is used for our particular model is described in 
\cite{s:wiseman}.

Finally, a different detector configuration may be chosen.  For example, a
drastically different observation, known as photon counting, gives rise to 
a Poisson (jump) process.  We refer to \cite{s:barchielli} for a full 
account of the quantum stochastic approach to observations in quantum 
optics.

\section{Geometry and dynamics of the filter}
\label{sec:systems}

In the previous section we introduced our physical model.  A detailed 
analysis resulted in the filtering equation (\ref{eq:qfilt}), where 
$\pi_t(X)$ is the best estimate of the observable $X$ given the 
observations $Y_{s\le t}$.  We will now study this equation in detail.

Note that (\ref{eq:qfilt}) is driven by the observation $Y_t$, which is 
a classical stochastic process.  Hence (\ref{eq:qfilt}) is entirely 
equivalent to a classical It\^o equation.  This is an important point, as 
it means that in the remainder of this article we only need classical 
stochastic calculus.

\subsection{The state space}

We begin by investigating the state space on which the filter evolves.  
Clearly (\ref{eq:qfilt}) defines the time evolution of a map $\pi_t$; we 
will show how this map can be represented efficiently.

The map $\pi_t$ associates to every observable $X$ on $\mathcal{H}$ a 
classical stochastic process which represents the expectation of $X$ 
conditioned on the observations up to time $t$.  It is easily verified 
that $\pi_t$ is linear, identity-preserving, and maps positive observables 
to positive numbers: in fact, it acts exactly like the expectation of $X$ 
with respect to some finite-dimensional state on $\mathcal{H}$.  We will 
denote this state by $\rho_t$, the conditional density at time $t$, where 
by definition $\pi_t(X)={\rm Tr}[\rho_tX]$.

It is straightforward to find an expression for $\rho_t$.  We get
\begin{equation}
\label{eq:qtraj}
	d\rho_t=\mathcal{L}^*\rho_t\,dt+\sqrt{\eta}(L\rho_t+\rho_tL^*-
		{\rm Tr}[\rho_t(L+L^*)]\rho_t)\,dW_t
\end{equation}
with the innovations $dW_t=dY_t-\sqrt{\eta}{\rm Tr}[\rho_t(L+L^*)]\,dt$ 
and the adjoint generator
$\mathcal{L}^*\rho=-i[H,\rho]+L\rho L^*-\tfrac{1}{2}(L^* 
L\rho+\rho L^* L)$.  In physics this equation is also known as a 
quantum trajectory equation or stochastic master equation.

Let $\dim\mathcal{H}=n$; as $n$ is finite, we can represent linear 
operators on $\mathcal{H}$ by complex matrices.  Thus (\ref{eq:qtraj})
is an ordinary, finite-dimensional It\^o equation.  We saw in section 
\ref{sec:finite} that $\rho_t$ is a density matrix, i.e.\ it belongs to 
the space
\begin{equation}
	\mathcal{P}=\{\rho\in\mathbb{C}^{n\times n}:\rho=\rho^*,
		~{\rm Tr}\rho=1,~\rho\ge 0\}.
\end{equation}
By construction $\mathcal{P}$ is an invariant set of (\ref{eq:qtraj}), and 
forms the natural state space of the filter.

\subsection{Geometry of $\mathcal{P}$}
\label{sec:kimura}

The geometry of $\mathcal{P}$ is rather complicated \cite{s:topology2}.  
To make the space more manageable we will reparametrize $\mathcal{P}$ so 
it can be expressed as a semialgebraic set.

Let us choose the matrix elements $\rho_{ij}$ of $\rho$ as follows.  For 
$i>j$ set $\rho_{ij}=\lambda_{ij}+i\mu_{ij}$ with $\lambda_{ij},
\mu_{ij}\in\mathbb{R}$.  For $i<j$ set $\rho_{ij}=\rho_{ji}^*$.  Finally, 
choose an integer $k$ between $1$ and $n$.  For $i\ne k$ set
$\rho_{ii}=\nu_i$, $\nu_i\in\mathbb{R}$, and $\rho_{kk}=1-\sum_{i\ne k}
\nu_i$.  Collect all $n^2-1$ numbers $\lambda_{ij},\mu_{ij},\nu_i$ into a 
vector $\Lambda$.  Then clearly the map $h:\Lambda\mapsto\rho$ is an 
isomorphism between $\mathbb{R}^{n^2-1}$ and 
$\{\rho\in\mathbb{C}^{n\times n}:\rho=\rho^*,~{\rm Tr}\rho=1\}$.

It remains to find the subset $K\subset\mathbb{R}^{n^2-1}$ that 
corresponds to positive definite matrices.  This is nontrivial, however, 
as it requires us to express nonnegativity of the eigenvalues of $\rho$ as 
constraints on $\rho_{ij}$.  The problem was solved by Kimura 
\cite{s:topology2} using Descartes' sign rule and the Newton-Girard 
identities for symmetric polynomials; we quote the following result:
\begin{proposition}
	Define $k_p(\rho)$, $p=2\ldots n$ recursively by
	\begin{equation}
		pk_p(\rho)=\sum_{q=1}^p(-1)^{q-1}\,{\rm Tr}[\rho^q]\,
		k_{p-q}(\rho)
	\end{equation}
	with $k_0=k_1=1$.  Define the semialgebraic set
	\begin{equation}
		K=\{ \Lambda\in\mathbb{R}^{n^2-1}:
			k_p(h(\Lambda))\ge 0,~p=2\ldots n \}.
	\end{equation}
	Then $h$ is an isomorphism between $K$ and $\mathcal{P}$.
\end{proposition}

Note that $2k_2=1-{\rm Tr}[\rho^2]\ge 0$ implies $\|\Lambda\|^2=
\sum_i\nu_i^2+\sum_{i>j}(\lambda_{ij}^2+\mu_{ij}^2) \le {\rm Tr}[h(\Lambda)^2]
\le 1$.  Hence $K$ is compact.

We work out explicitly the simplest case $n=2$ (spin $j=\tfrac{1}{2}$).  
Set $\rho_{11}=\nu$, $\rho_{22}=1-\nu$, $\rho_{21}=\lambda+i\mu=\rho_{12}^*$.
Then
\begin{equation}
	K_2=\{ \Lambda=(\lambda,\mu,\nu)\in\mathbb{R}^3:
		\lambda^2+\mu^2+\nu(\nu-1)\le 0
	\}.
\end{equation}
This is just a solid sphere with radius $\tfrac{1}{2}$, centered at 
$(0,0,\tfrac{1}{2})$.  The case $n=2$ is deceptively simple, however: 
it is the only case with a simple topology \cite{s:topology,s:topology2}.

We can also express (\ref{eq:qtraj}) in terms of $\Lambda$.  Specifically, 
we will consider the spin system $L=\sqrt{M}\,J_z$, $H=B(t)\,J_y$ in the 
basis $\psi_{1/2}=(1,0)$, $\psi_{-1/2}=(0,1)$ on $\mathbb{C}^{2\times 2}$.
We obtain
\begin{equation}
\label{eq:bloch}
\begin{split}
	d\lambda_t &= (B(t)(\nu_t-\tfrac{1}{2})-\tfrac{1}{2}M\lambda_t)\,dt+\sqrt{M\eta}\,\lambda_t(1-2\nu_t)\,dW_t \\
	d\mu_t     &= -\tfrac{1}{2}M\mu_t\,dt+\sqrt{M\eta}\,\mu_t(1-2\nu_t)\,dW_t \\
	d\nu_t     &= -B(t)\lambda_t\,dt-2\sqrt{M\eta}\,\nu_t(\nu_t-1)\,dW_t.
\end{split}
\raisetag{12pt}
\end{equation}
By construction, $K_2$ is an invariant set for this system.

\subsection{Convexity and pure states}

Just like its classical counterpart, the set of densities $\mathcal{P}$ is 
convex.  We have the following fundamental result:
\begin{proposition}
	The set $\mathcal{P}$ is the convex hull of the set of pure 
	states $\mathcal{Q}=\{vv^*\in\mathbb{C}^{n\times n}:
	v\in\mathbb{C}^n,~\|v\|=1\}\subset\mathcal{P}$.
\end{proposition}
\begin{proof}
	As any $\rho\in\mathcal{P}$ is self-adjoint it can be written as
	$\rho=\sum_i\lambda_iv_iv_i^*$, where $v_i$ are 
	orthonormal eigenvectors of $\rho$ and $\lambda_i$ are the 
	corresponding eigenvalues.  But ${\rm Tr}\rho=1,~\rho\ge 0$ 
	imply that $\sum_i\lambda_i=1$ and $\lambda_i\in[0,1]$.
	Hence $\mathcal{P}\subset{\rm conv}\,\mathcal{Q}$.  Conversely, it 
	is easily verified that ${\rm conv}\,\mathcal{Q}\subset\mathcal{P}$.
\end{proof}

Pure states are the extremal elements of $\mathcal{P}$; they represent 
quantum states of maximal information.  Note that classically extremal 
measures are deterministic, i.e.\ $\mathbb{P}[A]$ is either $0$ 
or $1$ for any event $A$.  This is not the case for pure 
states $\rho=vv^*$, however: any event $A=ww^*$ with $0<\|w^* 
v\|<1$, $\|w\|=1$ will have $0<{\rm Tr}[\rho A]<1$.  Thus no quantum state is 
deterministic, unless we restrict to a commutative algebra $\mathcal{A}$.

Intuitively one would expect that if the output $Y_t$ is not corrupted by 
independent noise, i.e.\ $\eta=1$, then there is no loss of information,
and hence an initially pure $\rho_0$ would remain pure under (\ref{eq:qtraj}).
This is indeed the case.  Define
\begin{equation}
\label{eq:statediff}
	dv_t=[
	(h_tL-\tfrac{1}{2}L^* L-\tfrac{1}{2}h_t^2-iH)\,dt +
	(L-h_t)\,dW_t
	]\,v_t
\end{equation}
where $h_t=\tfrac{1}{2}v_t^*(L^*+L)v_t$.  Then it is easily verified 
that $\rho_t=v_tv_t^*$ obeys (\ref{eq:qtraj}) with $\eta=1$.  It 
follows that if $\eta=1$, $\mathcal{Q}$ is an invariant set of 
(\ref{eq:qtraj}).  In the concrete example (\ref{eq:bloch}) it is not 
difficult to verify this property directly: when $\eta=1$, the sphere 
$\lambda^2+\mu^2+\nu(\nu-1)=0$ is invariant under (\ref{eq:bloch}).

\subsection{Quantum state reduction}
\label{sec:reduction}

We now study the dynamics of the spin filtering equation without feedback 
$B(t)=0$.  We follow the approach of \cite{s:adler}.

Consider the quantity $V_t=\pi_t(J_z^2)-\pi_t(J_z)^2$. We obtain
\begin{equation}
	\frac{d\,\mathbb{E}[V_t]}{dt}=
	-4M\eta\,\mathbb{E}[V_t^2].
\end{equation}
Clearly $\mathbb{E}[V_t^2]\ge 0$, so $\mathbb{E}[V_t]$ decreases 
monotonically.  But $V_t\ge 0$ and $\mathbb{E}[V_t^2]=0~{\rm 
iff}~V_t=0~{\rm a.s.}$  We conclude that
\begin{equation}
	\lim_{t\to\infty}\mathbb{E}[V_t]=0
\end{equation}
and hence $V_t\to 0~{\rm a.s.}$\ as $t\to\infty$.  But the only states 
$\rho\in\mathcal{P}$ with $V_t={\rm Tr}[J_z^2\rho]-{\rm Tr}[J_z\rho]^2=0$
are the eigenstates $\psi_m\psi_m^*$ of $J_z$.  Hence in the long-time 
limit the conditional state {\it collapses} onto one of the eigenstates of 
$J_z$, as predicted by (\ref{eq:collapse}) for a ``direct'' measurement of 
$J_z$.

With what probability does the state collapse onto eigenstate $m$?
To study this, let us calculate $\pi_t(\psi_m\psi_m^*)$.  We get
\begin{equation}
	d\pi_t(\psi_m\psi_m^*)=
	2\sqrt{M\eta}\,\pi_t(\psi_m\psi_m^*)(m-\pi_t(J_z))\,dW_t.
\end{equation}
Clearly $\pi_t(\psi_m\psi_m^*)$ is a martingale, so
\begin{equation}
\label{eq:collpro}
	p_m=\mathbb{E}[\pi_\infty(\psi_m\psi_m^*)]
		=\pi_0(\psi_m\psi_m^*).
\end{equation}
We have already shown that $\rho_\infty$ is one of $\psi_n\psi_n^*$, 
and as the $\psi_m$ are orthonormal this implies that 
$\pi_\infty(\psi_m\psi_m^*)={\rm Tr}[\rho_\infty\psi_m\psi_m^*]$
is $1$ if $n=m$ and $0$ otherwise.  Thus $p_m$ is just the probability of 
collapsing onto the eigenstate $m$.  But note that 
$\pi_0(\psi_m\psi_m^*)={\rm Tr}[\rho_0\psi_m\psi_m^*]$, so
(\ref{eq:collpro}) gives exactly the same collapse probability as the 
``direct'' measurement (\ref{eq:collprodir}).

We conclude that the predictions of quantum filtering theory are entirely 
consistent with the traditional quantum mechanics.  A continuous reduction 
process replaces, but is asymptotically equivalent to, the instantaneous 
state collapse of section \ref{sec:finite}.  This phenomenon is known as 
{\it quantum state reduction}\footnote{
	The term state reduction is sometimes associated with quantum
	state diffusion, an attempt to empirically modify the laws of
	quantum mechanics so that state collapse becomes a dynamical 
	property.  The state diffusion equation, which is postulated
	rather than derived, is exactly (\ref{eq:statediff}) with $L=L^*$.
	We use the term state reduction as describing the reduction 
	dynamics without any relation to its interpretation.
	The analysis of Ref.\ \cite{s:adler} is presented in the context
	of quantum state diffusion, but applies equally well to our case.
}.  We emphasize that quantum filtering is purely a statistical inference 
process and is obtained entirely through nondemolition measurements.
Note also that state reduction occurs because $L=J_z$ is self-adjoint; 
other cases are of equal physical interest, but we will not consider them 
in this paper.

Physically, the filtering approach shows that realistic measurements are 
not instantaneous but take some finite time.  The time scale of state 
reduction is of order $M^{-1}$, an experimentally controlled parameter.
A carefully designed experiment can thus have a reduction time scale of an
order attainable by modern digital electronics \cite{s:FPGA}, which opens 
the door to both measuring and manipulating the process in real time.

\section{Stabilization of spin state reduction}
\label{sec:spin}

\begin{figure}
\centering
\includegraphics[width=3in]{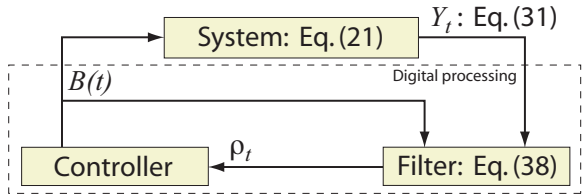}
\caption{Schematic of the feedback control strategy.  The output from the 
system is used to propagate the conditional state of the filter.  The 
feedback signal is of state feedback form with respect to the conditional 
state.}
\label{f:cschematic}
\end{figure}

\subsection{The control problem}

It is a standard idea in stochastic control that an output feedback 
control problem can be converted into a state feedback problem for the 
filter \cite{s:mortensen,s:astrom}.  This is shown schematically in Fig.\ 
\ref{f:cschematic}. The filtering equations (\ref{eq:qfilt}) or
(\ref{eq:qtraj}) are driven by $Y_t$; hence, at least in principle, the 
conditional state $\rho_t$ can be calculated recursively in real time by a 
digital processor.  

The filter describes optimally our knowledge of the system; clearly the 
extent of our knowledge of the system state limits the precision with 
which it can be controlled.  The best we can hope to do is to control the 
system to the best of our knowledge, i.e.\ to control the filter.  The 
latter is a well-posed problem, despite that we cannot predict the 
observations $Y_t$, because we know the statistics of the innovations 
process $W_t$.

For such a scheme to be successful the system dynamics (\ref{eq:qito}) 
must be known, as the optimal filter is matched to the system dynamics.  
Designing controllers that perform well even when the system dynamics is 
not known precisely is the subject of robust control theory.  Also, 
efficient signal processing algorithms and hardware are necessary 
to propagate (\ref{eq:qtraj}) in real time, which is particularly 
problematic when $\dim\mathcal{H}$ is large.  Neither of these issues will 
be considered in this paper.

The state reduction dynamics discussed in the previous section immediately 
suggests the following control problem: we wish to find state feedback 
$B(t)=\Phi(\rho_t)$ so that one of the eigenstates $\rho=\psi_m\psi_m^*$
is globally stabilized.  The idea that a quantum measurement can be 
engineered to collapse deterministically onto an eigenstate of our choice 
is somewhat remarkable from a traditional physics perspective, but clearly 
the measurement scenario we have described provides us with this opportunity.
For additional motivation and numerical simulations relating to this 
control problem, see \cite{s:stockton}.


\begin{figure*}
\centering
\includegraphics[width=\textwidth]{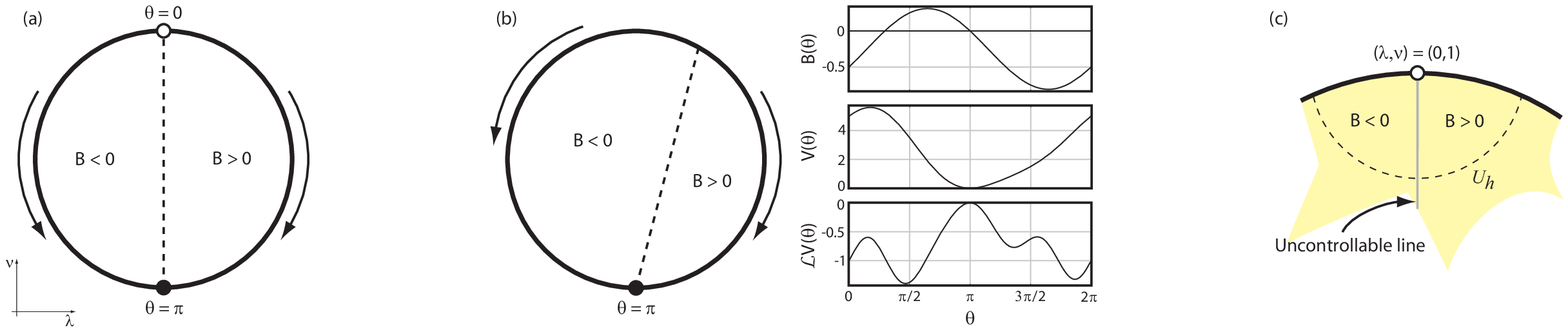}
\caption{Cartoons of the various control schemes; the arrows denote the 
	rotation direction of the magnetic field. (a) Almost global control 
	on the circle: the magnetic field always rotates in the direction 
	of least distance to $\theta=\pi$, but $\theta=0$ remains a fixed 
	point. (b) Global control on the circle: we intentionally break 
	the symmetry of the controller to remove the undesired fixed 
	point.  The graphs show a typical feedback law and Lyapunov design 
	with $M=1$, $B(\theta)=\tfrac{1}{2}\sin\theta-\tfrac{1}{4}(1+\cos\theta)$, 
	$V(\theta)=(\tfrac{5}{2}+\sin\theta)(1+\cos\theta)$.
	(c) A neighborhood of $(\lambda,\nu)=(0,1)$ showing why 
	the almost global control law fails on the disc.  The control 
	vanishes on the line $\lambda=0$; hence points on this line are 
	never repelled with unit probability, in violation of (\ref{eq:repel}).}
\label{f:control}
\end{figure*}


\subsection{Stochastic stability}
\label{sec:stab}

In nonlinear control theory \cite{s:nijmeijer} stabilization of nonlinear 
systems is usually performed using the powerful tools of Lyapunov 
stability theory.  In this section we will describe the stochastic 
counterpart of deterministic Lyapunov theory, developed in the 1960s by 
Has'minski\u{\i} and others.  We will not give proofs, for which we refer 
to \cite{s:hasminskii,s:kushner,s:arnold,s:mao}.

Let $W_t$ be a Wiener process on the canonical Wiener space 
$(\Omega,\mathcal{F},\mathbb{P})$.  Consider an It\^o equation on 
$\mathbb{R}^n$ of the form
\begin{equation}
\label{eq:SDE}
	dx_t=b(x_t)\,dt+\sigma(x_t)\,dW_t
\end{equation}
where $b,\sigma:\mathbb{R}^n\to\mathbb{R}^n$ satisfy the usual linear 
growth and local Lipschitz conditions for existence and uniqueness of 
solutions \cite{s:rogersw}.  Let $x^*$ be a fixed point of (\ref{eq:SDE}), 
i.e.\ $b(x^*)=\sigma(x^*)=0$.
\begin{definition}
The equilibrium solution $x_t=x^*$ of (\ref{eq:SDE}) is
\begin{enumerate}
\item	{\it stable in probability} if
	\begin{equation*}
		\lim_{x_0\to x^*}\mathbb{P}\left[
			\sup_{t\ge 0}|x_t-x^*|>\epsilon
		\right]=0~~~~~~~\forall\epsilon>0,
	\end{equation*}
\item	{\it asymptotically stable} if it is stable in probability and
	\begin{equation*}
		\lim_{x_0\to x^*}\mathbb{P}\left[
			\lim_{t\to\infty}|x_t-x^*|=0
		\right]=1,
	\end{equation*}
\item	{\it globally stable} if it is stable in probability and
	\begin{equation*}
		\mathbb{P}\left[
			\lim_{t\to\infty}|x_t-x^*|=0
		\right]=1.
	\end{equation*}
\end{enumerate}
Note that 1 and 2 are local properties, whereas 3 is a global property of 
the system.
\end{definition}

Recall that the infinitesimal generator of $x_t$ is given by
\begin{equation}
	\mathscr{L}=\sum_i b^i(x)\frac{\partial}{\partial x^i}+
	\frac{1}{2}\sum_{ij}\sigma^i(x)\sigma^j(x)
		\frac{\partial^2}{\partial x^i\partial x^j}
\end{equation}
so $d\,\mathbb{E}[f(x_t)]/dt=\mathbb{E}[\mathscr{L}f(x_t)]$.  We can 
now state the stochastic equivalent of Lyapunov's direct method 
\cite{s:hasminskii,s:kushner,s:arnold}.
\begin{theorem}
\label{th:lyapunov}
	Define $U_h=\{x:|x-x^*|<h\}$.
	Suppose there exists some $h>0$ and a function $V:U_h\to\mathbb{R}_+$ 
	that is continuous and twice differentiable on $U_h\backslash\{x^*\}$,
	such that $V(x^*)=0$ and $V(x)>0$ otherwise, and
	$\mathscr{L}V(x)\le 0$ on $U_h$.  Then the equilibrium solution
	$x_t=x^*$ is stable in probability.  If $\mathscr{L}V(x)<0$ on
	$U_h\backslash\{x^*\}$ then $x^*$ is asymptotically stable.
\end{theorem}

Theorem \ref{th:lyapunov} is a local theorem; to prove global stability 
we need additional methods.  When dealing with quantum filtering equations 
a useful global result is the following stochastic LaSalle-type theorem 
of Mao \cite{s:mao}.  In the theorem we will assume that the dynamics of 
(\ref{eq:SDE}) are confined to a {\it bounded invariant set} $G$.  
\begin{theorem}
\label{th:lasalle}
	Let $G$ be a bounded invariant set with respect to the solutions
	of (\ref{eq:SDE}) and $x_0\in G$.  Suppose there exists a
	continuous, twice differentiable function $V:G\to\mathbb{R}_+$ 
	such that $\mathscr{L}V(x)\le 0$ $\forall x\in G$.  Then 
	$\lim_{t\to\infty}\mathscr{L}V(x_t)=0~{\rm a.s.}$
\end{theorem}

Finally, we will find it useful to prove that a particular fixed point 
{\it repels} trajectories that do not originate on it.  To this end we use 
the following theorem of Has'minski\u{\i} \cite{s:hasminskii}.
\begin{theorem}
\label{th:hasminskii}
	Suppose there exists some $h>0$ and a function $V:U_h\to\mathbb{R}$ 
	that is continuous and twice differentiable on $U_h\backslash\{x^*\}$,
	such that
	\begin{equation*}
		\lim_{x\to x^*}V(x)=+\infty
	\end{equation*}
	and $\mathscr{L}V(x)<0$ on $U_h\backslash\{x^*\}$.  Then the 
	equilibrium solution $x_t=x^*$ is not stable in probability, and
	moreover
	\begin{equation}
	\label{eq:repel}
		\mathbb{P}\left[
			\sup_{t>0}|x_t-x^*|<h
		\right]=0
		~~~~~~~\forall x_0\in U_h\backslash\{x^*\}.
	\end{equation}
\end{theorem}

\subsection{A toy problem: the disc and the circle}

We treat in detail an important toy problem: spin $j=\tfrac{1}{2}$.  The 
low dimension and the simple topology make this problem easy to visualize.  
Nonetheless we will see that the stabilization problem is not easy to 
solve even in this simple case.

We have already obtained the filter (\ref{eq:bloch}) on $K_2$ for this 
case.  Conveniently, the origin in $K_2$ is mapped to the lower eigenstate 
$\psi_{-1/2}\psi_{-1/2}^*$; we will attempt to stabilize this state.

Note that the equations for $\lambda_t,\nu_t$ are decoupled from $\mu_t$.
Moreover, the only point in $K_2$ with $(\lambda,\nu)=(0,0)$ has $\mu=0$.
Hence we can equivalently consider the control problem
\begin{equation}
\label{eq:disc-ctl}
\begin{split}
	d\lambda_t &= (B(t)(\nu_t-\tfrac{1}{2})-\tfrac{1}{2}M\lambda_t)\,dt+\sqrt{M\eta}\,\lambda_t(1-2\nu_t)\,dW_t \\
	d\nu_t     &= -B(t)\lambda_t\,dt-2\sqrt{M\eta}\,\nu_t(\nu_t-1)\,dW_t
\end{split}
\raisetag{12pt}
\end{equation}
on the disc $B^2=\{(\lambda,\nu)\in\mathbb{R}^2:\lambda^2+\nu(\nu-1)\le 0\}$.
Controlling (\ref{eq:disc-ctl}) is entirely equivalent to controlling 
(\ref{eq:bloch}), as globally stabilizing $(\lambda,\nu)=(0,0)$ guarantees
that $\mu$ is attracted to zero due to the geometry of $K_2$.

An even simpler toy problem is obtained as follows.  Suppose $\eta=1$; we 
have seen that then the sphere $\lambda^2+\mu^2+\nu(\nu-1)=0$ is invariant 
under (\ref{eq:bloch}).  Now suppose that additionally $\mu_0=0$.  Then 
clearly the circle $S^1=\{(\lambda,\nu)\in\mathbb{R}^2:
\lambda^2+\nu(\nu-1)=0\}$ is an invariant set.  We find
\begin{equation}
\label{eq:circlec}
	d\theta_t=(B(t)-\tfrac{1}{2}M\sin\theta_t\cos\theta_t)\,dt
	-\sqrt{M}\sin\theta_t\,dW_t
\end{equation}
after a change of variables $(2\lambda_t,2\nu_t)=(\sin\theta_t,1+\cos\theta_t)$.

The system (\ref{eq:disc-ctl}) could in principle be realized by 
performing the experiment of Fig.\ \ref{f:schematic} with a single atom.
The reduced system (\ref{eq:circlec}) is unrealistic, however: it would 
require perfect photodetectors and perfect preparation of the initial 
state.  Nonetheless it is instructive to study this case, as it provides 
intuition which can be applied in more complicated scenarios.  Note that 
(\ref{eq:circlec}) is a special case of (\ref{eq:disc-ctl}) where $\eta=1$ 
and the dynamics is restricted to the boundary of $B^2$.

\subsection{Almost global control on $S^1$}

We wish to stabilize $(\lambda,\nu)=(0,0)$, which corresponds to 
$\theta=\pi$.  Note that by (\ref{eq:circlec}) a positive magnetic field 
$B>0$ causes an increasing drift in $\theta$, i.e.\ a clockwise rotation 
on the circle.  Hence a natural choice of controller is one which causes 
the state to rotate in the direction nearest to $\theta=\pi$ from the 
current position.  This situation is sketched in Fig.\ \ref{f:control}a.

A drawback of any such controller is that by symmetry, the feedback must 
vanish not only on $\theta=\pi$ but also on $\theta=0$; hence $\theta=0$ 
remains a fixed point of the controlled system and the system is not 
{\it globally} stable.  We will show, however, that under certain 
conditions such feedback renders the system {\it almost globally} stable,
in the sense that all paths that do not start on $\theta=0$ are attracted 
to $\theta=\pi$ $\rm a.s.$

For simplicity we choose a controller that is linear in $(\lambda,\nu)$:
\begin{equation}
\label{eq:semiglobal}
	B(t)=2G\lambda_t=G\sin\theta_t,~~~~~~~G>0.
\end{equation}
Here $G$ is the feedback gain.  The generator of (\ref{eq:circlec}) is then
\begin{equation}
	\mathscr{L}=(G\sin\theta-\tfrac{1}{2}M\sin\theta\cos\theta)
		\frac{\partial}{\partial\theta}+
	\tfrac{1}{2}M\sin^2\theta\,\frac{\partial^2}{\partial\theta^2}.
\end{equation}
As a first step we will show that the fixed point $\theta=\pi$ is 
asymptotically stable and that the system is always attracted to one of 
the fixed points (there are no limit cycles etc.)
To this end, consider the Lyapunov function
\begin{equation}
	V(\theta)=1+\cos\theta,~~~~~~~
		V(\pi)=0,~V(\theta\ne\pi)>0.
\end{equation}
We obtain
\begin{equation}
	\mathscr{L}V(\theta)=-G\sin^2\theta.
\end{equation}
It follows from Theorem \ref{th:lyapunov} that $\theta=\pi$ is 
asymptotically stable, and from Theorem \ref{th:lasalle} that 
$\lim_{t\to\infty}\theta_t\in\{0,\pi\}~{\rm a.s.}$

What remains to be shown is that any trajectory which does not start on 
$\theta=0$ ends up at $\theta=\pi~{\rm a.s.}$
To prove this, consider
\begin{equation}
	\tilde V(\theta)=-\log(1-\cos\theta),~~~~~~~
	\lim_{\theta\to 0}\tilde V(\theta)=+\infty.
\end{equation}
We easily find
\begin{equation}
	\mathscr{L}\tilde V(\theta)=
	\cos^2(\theta/2)\,(M+M\cos\theta-2G).
\end{equation}
Now note that
\begin{equation}
	\mathscr{L}\tilde V(\theta)<0~~
		\forall\,\theta\in(-\pi,\pi)\backslash\{0\}
	~~~{\rm iff}~~~G\ge M.
\end{equation}
Thus by Theorem \ref{th:hasminskii} we have
\begin{equation}
	\mathbb{P}\left[
		\sup_{t>0}|\theta_t|<\pi
	\right]=0~~~{\rm if}~~~\theta_0\in(-\pi,\pi)\backslash\{0\}.
\end{equation}
But as $\theta\in S^1$ this implies $\theta_t\to\pi~{\rm a.s.}$ if
$\theta_0\in(-\pi,\pi)\backslash\{0\}$.  We conclude that the control law
(\ref{eq:semiglobal}) almost globally stabilizes the system if we have 
sufficient gain $G\ge M$.

\subsection{Global control on $S^1$}

Any deterministic system on the circle is topologically obstructed\footnote{
	This is only the case for systems with continuous vector fields 
	and continuous, pure state feedback.  The obstruction can be 
	lifted if one considers feedback laws that are discontinuous or 
	that have explicit time dependence.
} from having a globally stabilizing controller: a continuous vector field 
on $S^1$ with a stable fixed point necessarily has an unstable fixed point 
as well.  In the stochastic case, however, this is not the case.  Though 
the drift and diffusion terms must each have two fixed points, we may 
design the system in such a way that only the stable fixed points coincide.

To apply such a trick in our system we must break the natural symmetry 
of the control law.  This situation is shown in Fig.\ \ref{f:control}b.
There is a region of the circle where the control rotates in the direction 
with a longer distance to $\theta=\pi$; the advantage is that $\theta=0$ 
is no longer a fixed point.

The linear control law that has this property has the form
\begin{equation}
	B(t)=2G\lambda_t+2H\nu_t=G\sin\theta_t+H(1+\cos\theta_t)
\end{equation}
with $G>0$.  We can prove global stability by applying Theorems 
\ref{th:lyapunov} and \ref{th:lasalle} with a Lyapunov function of the form
\begin{equation}
	V(\theta)=(\alpha+\sin\theta)(1+\cos\theta),~~~~~~~\alpha>1.
\end{equation}
Unfortunately it is not obvious from the analytic form of $\mathscr{L}V$ 
how $\alpha$ must be chosen to satisfy the Lyapunov condition.  It is 
however straightforward to plot $\mathscr{L}V$, so that in this simple
case it is not difficult to search for $\alpha$ by hand.

A typical design for a particular choice of parameters is shown in Fig.\ 
\ref{f:control}b.  The conditions of Theorems \ref{th:lyapunov} and 
\ref{th:lasalle} are clearly satisfied, proving that the system is 
globally stable.  Note that when the symmetry is broken we no longer need 
to fight the attraction of the undesired fixed point; hence there is no 
lower bound on $G$.  In fact, in Fig.\ \ref{f:control}b we have $G<M$.

\subsection{Almost global control on $B^2$}

Unfortunately, the simple almost global control design on $S^1$ does not 
generalize to $B^2$.  The problem is illustrated in Fig.\ \ref{f:control}c.
The controller (\ref{eq:semiglobal}) vanishes at $\theta=0$ and $\pi$, but 
we can prove that $\theta=0$ is repelling.  On $B^2$, however, the control 
vanishes on the entire line $\lambda=0$ which becomes an invariant set of 
(\ref{eq:disc-ctl}).  But then it follows from (\ref{eq:collpro}) that any 
trajectory with $\lambda_0=0$, $\nu_0\not\in\{0,1\}$ has a nonzero 
probability of being attracted to either fixed point.

Consider a neighborhood $U_h$ of the point $(\lambda,\nu)=(0,1)$
that we wish to destabilize.  For any $h>0$, however small, $U_h$
contains points on the line $\lambda=0$ for which $\nu<1$, and 
we have seen that trajectories starting at such points have a nonzero 
probability of being attracted to $(0,1)$.  But this violates 
(\ref{eq:repel}), so clearly we cannot prove Theorem \ref{th:hasminskii} 
on $B^2$.

One could attempt to prove that all points except those with $\lambda=0$ 
are attracted to the origin with unit probability.  The Lyapunov theory of 
section \ref{sec:stab} is not equipped to handle such a case, however, and 
new methods must be developed \cite{s:vanhandel}.  Instead, we will focus 
on the global control problem.

\subsection{Global control on $B^2$ and semialgebraic geometry}

Once again we consider the asymmetric control law
\begin{equation}
	B(t)=2G\lambda_t+2H\nu_t,~~~~~~~G>0
\end{equation}
and try to show that it globally stabilizes the system.  Before we can 
solve this problem, however, we must find a systematic method for proving 
global stability.  Searching ``by hand'' for Lyapunov functions is clearly 
impractical in two dimensions, and is essentially impossible in higher 
dimensions where the state space cannot be visualized.  

In fact, even if we are given a Lyapunov function $V$, testing whether 
$\mathscr{L}V\le 0$ on $K$ is highly nontrivial.  The problem can be 
reduced to the following question: is the set $\{\Lambda\in\mathbb{R}^{n^2-1}:
\mathscr{L}V>0,~k_p(h(\Lambda))\ge 0,~p=2\ldots n\}$ empty?
Such problems are notoriously difficult to solve and their solution is 
known to be NP-hard in general \cite{s:parrilo}.

The following result, due to Putinar \cite{s:putinar}, suggests one way to 
proceed.  Let $S$ be a semialgebraic set, i.e.\ 
$S=\{x\in\mathbb{R}^m:s_i(x)\ge 0,~i=1\ldots n\}$ with polynomial $s_i$.
Suppose that for some $i$ the set $\{x\in\mathbb{R}^m:s_i(x)\ge 0\}$ is 
compact.  Then any polynomial $p$ that is strictly positive on $S$ is of 
the form
\begin{equation}
\label{eq:putinar}
	p(x)=p_0(x)+\sum_{i=1}^np_i(x)s_i(x),~~
	p_k(x)=\sum_j p_{kj}(x)^2
\end{equation}
where $p_{kj}$ are polynomials; i.e., $p$ is an affine combination of 
the constraints $s_i$ and {\it sum-of-squares} polynomials $p_k$.

Conversely, it is easy to check that any polynomial of the form 
(\ref{eq:putinar}) is nonnegative on $S$.  We may thus consider the 
following relaxation: instead of testing nonnegativity of a polynomial on 
$S$, we may test whether the polynomial can be represented in the form 
(\ref{eq:putinar}).  Though it is not true that any nonnegative polynomial 
on $S$ can be represented in this form, Putinar's result suggests that the 
relaxation is not overly restrictive.  The principal advantage of this 
approach is that the relaxed problem can be solved in polynomial time 
using semidefinite programming techniques \cite{s:parrilothesis,s:prajna}.

The approach is easily adapted to our situation as $K$ is a semialgebraic
set, and we solve the relaxed problem of testing whether $-\mathscr{L}V$ 
can be expressed in the form (\ref{eq:putinar}).  In fact, the 
semidefinite programming approach of \cite{s:parrilothesis,s:prajna} even 
allows us to search for polynomial $V$ such that (\ref{eq:putinar}) is 
satisfied; hence we can search numerically for a global stability proof 
using a computer program.  Such searches are easily implemented using the 
Matlab toolbox SOSTOOLS \cite{s:sostools}.

\begin{figure}
\centering
\includegraphics[width=\linewidth]{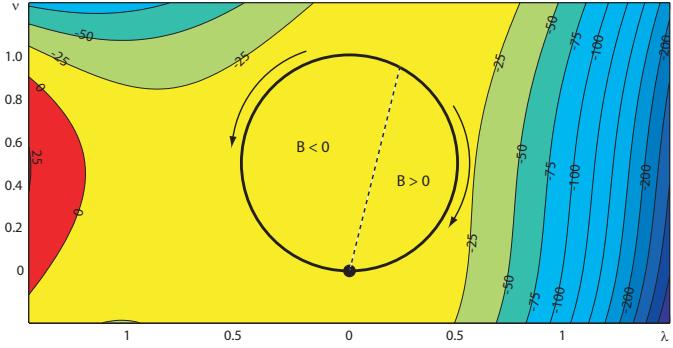}
\caption{Contour plot of $\mathscr{L}V$ for the control law
$B(t)=4\lambda_t-\nu_t$, with $M=2$ and $\eta=\tfrac{1}{2}$.  
The function $V$ was found by semidefinite programming.}
\label{f:sossoln}
\end{figure}

A typical design for a particular choice of parameters is shown in Fig.\ 
\ref{f:sossoln}.  After fixing the parameters $M=2$, $\eta=\tfrac{1}{2}$, 
and the control law $B(t)=4\lambda_t-\nu_t$, an SOSTOOLS search found the 
Lyapunov function
\begin{equation}
	V(\lambda,\nu)=
	21.8\,\nu-5.73\,\lambda^2+10.4\,\lambda\nu-5.63\,\nu^2
\end{equation}
where $-\mathscr{L}V$ is of the form (\ref{eq:putinar}).  Hence Theorems 
\ref{th:lyapunov} and \ref{th:lasalle} are satisfied, proving that the 
system is globally stable.

A couple of technical points should be made at this point.  Note that 
formally the filtering equation (\ref{eq:qtraj}) and its parametrizations 
do not satisfy the linear growth condition.  However, as the filter
evolves on a compact invariant set $K$, we could modify the equations 
smoothly outside $K$ to be of linear growth without affecting the dynamics 
in $K$.  Hence the results of section \ref{sec:stab} can still be used.  
Moreover, it is also not strictly necessary that $V$ be nonnegative, as 
adding a constant to $V$ does not affect $\mathscr{L}V$.  Hence it is 
sufficient to search for polynomial $V$ using SOSTOOLS.

\subsection{Global control for higher spin}

The approach for proving global stability described in the previous 
section works for arbitrary spin $j$.  To generalize our control scheme 
we need to convert to the parametrization of section \ref{sec:kimura}, as 
we did for spin $j=\tfrac{1}{2}$ in (\ref{eq:disc-ctl}).  We must also 
propose a control law that works for general spin systems.

We do not explicitly convert to the parametrized form or generate the 
constraints $k_p$, as this procedure is easily automated using Matlab's 
symbolic toolbox.  Note that the parameter $k$ determines which eigenstate 
is mapped to the origin.  This is convenient for SOSTOOLS searches, as 
polynomials can be fixed to vanish at the origin simply by removing the 
constant term.  We always wish to stabilize the origin in the parametrized 
coordinate system.

To speed up computations we can eliminate all the parameters $\mu_{ij}$ as 
was done in going from (\ref{eq:bloch}) to (\ref{eq:disc-ctl}).  The fact 
that the remaining equations are decoupled from $\mu_{ij}$ is easily seen 
from (\ref{eq:qtraj}), as both $iJ_y$ and $J_z$ are real matrices.  
Moreover it is easily verified that, by convexity of $K$, the orthogonal  
projection of any $\rho\in K$ onto $\{\mathbb{R}^{n^2-1}:
\mu_{ij}=0~\forall i>j\}$ lies inside $K$.  Hence we only need to consider 
the reduced control problem with $\mu_{ij}=0$.

In \cite{s:stockton} we numerically studied two control laws for general 
spin systems.  The first law, $B_1(t)=\pi_t(J_xJ_z+J_zJ_x-2m_dJ_x)$ ($m_d$ 
is the eigenstate we wish to stabilize), reduces to our almost global 
control law when $j=\tfrac{1}{2}$.  However, numerical simulations suggest 
that for $j>\tfrac{1}{2}$ this control law gives a finite collapse 
probability onto $m\ne m_d$.  The second law, $B_2(t)=\pi_t(J_z)-m_d$, 
reduces to $B_2(t)=\nu_t$ in the case $j=\tfrac{1}{2}$, which is not 
locally stable.  Our experience with $j=\tfrac{1}{2}$ suggests that a 
control law of the form
\begin{equation}
\label{eq:generalc}
	B(t)=G\,\pi_t(J_xJ_z+J_zJ_x-2m_dJ_x)+H\,(\pi_t(J_z)-m_d)
\end{equation}
should globally stabilize the eigenstate $m_d$ of a spin $j$ system.

We have verified global stability for a typical design with $j=1$, $M=2$, 
$\eta=\tfrac{1}{2}$, and $B(t)=2\,\pi_t(J_xJ_z+J_zJ_x)+\pi_t(J_z)$ using 
SOSTOOLS.  A Lyapunov function was indeed found that guarantees global 
stability of the eigenstate $\psi_0\psi_0^*$.

Physically the case $j>\tfrac{1}{2}$ is much more interesting than 
$j=\tfrac{1}{2}$.  An experiment with $j>\tfrac{1}{2}$ can be performed 
with multiple atoms, in which case the control produces statistical 
correlations between the atoms.  Such correlations, known as entanglement, 
are important in quantum computing.  The structure of the control problem 
is, however, essentially the same for any $j$.  We refer to 
\cite{s:stockton,s:stockton2} for details on entanglement generation in 
spin systems.

\section{Conclusion}

In this paper we have argued that quantum mechanical systems that are 
subjected to measurement are naturally treated within the framework of
(albeit noncommutative) stochastic filtering theory.  The quantum control 
problem is then reduced to a classical stochastic control problem for the 
filter.  We have demonstrated the viability of this approach by 
stabilizing state reduction in simple quantum spin systems using 
techniques of stochastic nonlinear control theory.

Unfortunately, the stabilization techniques of section \ref{sec:spin} have 
many drawbacks.  We do not have a systematic procedure for finding control 
laws: we postulate linear controllers and search for corresponding 
Lyapunov functions.  Even when the control law is known, verifying 
global stability is nontrivial even in the simplest case.  Our numerical 
approach, though very successful in the examples we have shown, rapidly
becomes intractable as the dimension of the Hilbert space grows.  Finally, 
our methods do not allow us to make general statements; for example, 
though it seems plausible that the control law (\ref{eq:generalc}) is 
globally stabilizing for any $j,m_d,M,\eta$, $H\ne 0$ and $G>0$, we have 
not yet succeeded in proving such a statement.

Nonetheless we believe that the general approach outlined in this paper 
provides a useful framework for the control of quantum systems.  It is 
important in this context to develop methods for the control of classical 
stochastic nonlinear systems \cite{s:florch1,s:florch2,s:florch3,s:florch4},
as well as methods that exploit the specific structure of quantum control 
problems.  The design of realistic control systems will also require 
efficient signal processing algorithms for high-dimensional quantum 
filtering and methods for robust quantum control \cite{s:james}.

\section*{Acknowledgment}
The authors thank Luc Bouten, Andrew Doherty, Richard Murray and Stephen 
Prajna for enlightening discussions.

\bibliographystyle{IEEEtran}
\bibliography{IEEEabrv,stab-arxiv}

\end{document}